%% file: main.tex
\documentclass[sigconf, nonacm]{acmart}
\AtBeginDocument{%
  }

\copyrightyear{2025}
\acmYear{2025}
\setcopyright{acmlicensed}\acmConference[SIGSPATIAL '25]{The 33rd ACM International Conference on Advances in Geographic Information Systems}{November 3--6, 2025}{Minneapolis, MN, USA}
\acmBooktitle{The 33rd ACM International Conference on Advances in Geographic Information Systems (SIGSPATIAL '25), November 3--6, 2025, Minneapolis, MN, USA}
\acmDOI{10.1145/3748636.3764176}
\acmISBN{979-8-4007-2086-4/2025/11}

\usepackage{adjustbox}
\usepackage{subfigure}
\usepackage{multirow}
\usepackage{graphicx}
\usepackage{subcaption}

\newcommand{\modelname}{\textsc{CooKIE }}
\newcommand{\modelnamex}{\textsc{CooKIE}}
\newcommand{\modelnamelogo}{\textsc{CooKIE}\includegraphics[height=1em]{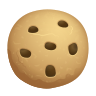}}

\newcommand{\predictor}{\textsc{Spatial-MGAT }}
\newcommand{\predictorx}{\textsc{Spatial-MGAT}}

\begin{document}

\title{Transit for All: Mapping Equitable Bike2Subway Connection using Region Representation Learning} 


\author{Min Namgung}
\email{namgu007@umn.edu}
\affiliation{%
  \institution{University of Minnesota, Twin Cities}
  \city{Minneapolis}
  \state{Minnesota}
  \country{USA}
}

\author{JangHyeon Lee}
\email{lee04588@umn.edu}
\affiliation{%
  \institution{University of Minnesota, Twin Cities}
  \city{Minneapolis}
  \state{Minnesota}
  \country{USA}
}

\author{Fangyi Ding}
\email{fyding@connect.hku.hk}
\affiliation{%
  \institution{The University of Hong Kong}
  \city{Hong Kong SAR}
  \country{China}
}

\author{Yao-Yi Chiang}
\email{yaoyi@umn.edu}
\affiliation{%
  \institution{University of Minnesota, Twin Cities}
  \city{Minneapolis}
  \state{Minnesota}
  \country{USA}
}

\renewcommand{\shortauthors}{Namgung et al.}

\begin{abstract}
Ensuring equitable public transit access remains challenging, particularly in densely populated cities like New York City (NYC), where low-income and minority communities often face limited transit accessibility. Bike-sharing systems (BSS) can bridge these equity gaps by providing affordable first- and last-mile connections. However, strategically expanding BSS into underserved neighborhoods is difficult due to uncertain bike-sharing demand at newly planned (``cold-start'') station locations and limitations in traditional accessibility metrics that may overlook realistic bike usage potential.
We introduce \textbf{Transit for All (TFA)}, a spatial computing framework designed to guide the equitable expansion of BSS through three components: (1) spatially-informed bike-sharing demand prediction at cold-start stations using region representation learning that integrates multimodal geospatial data, (2) comprehensive transit accessibility assessment leveraging our novel weighted Accessibility Level (wAL) by combining predicted bike-sharing demand with conventional transit accessibility metrics, and (3) strategic recommendations for new bike station placements that consider potential ridership and equity enhancement.
Using NYC as a case study, we identify transit accessibility gaps that disproportionately impact low-income and minority communities in historically underserved neighborhoods. Our results show that strategically placing new stations guided by wAL notably reduces disparities in transit access related to economic and demographic factors. From our study, we demonstrate that TFA provides practical guidance for urban planners to promote equitable transit and enhance the quality of life in underserved urban communities.

\end{abstract}





\maketitle
\input{section/introduction}

\input{section/data}
\input{section/related_works}

\input{section/method}
\input{section/experiment}
\input{section/result}

\input{section/conclusion}



\bibliographystyle{ACM-Reference-Format}
\bibliography{0_references}

\end{document}

%% file: section/introduction.tex
\section{Introduction}
Equitable access to public transit remains a critical global challenge due to uneven geographic distribution of transportation infrastructure, disproportionately affecting low-income and minority neighborhoods~\cite{goodman2014inequalities, ursaki2015quantifying}. Residents in these communities often experience long travel times and multiple transfers to reach transit hubs, limiting economic mobility~\cite{jiao2013transit} and reducing overall quality of life~\cite{owen2015modeling}. Addressing these inequities directly aligns with the UN Sustainable Development Goal (SDG) 11.2, which aims to provide accessible and sustainable transportation for all by 2030~\cite{un2030agenda}.

Bike-sharing systems (BSS) can mitigate these transit inequities by providing affordable first- and last-mile connections, enhancing urban connectivity~\cite{shaheen2010bikesharing, fishman2013bike}. BSS offer short-term bicycle rentals through distributed station networks, enabling convenient trips between transit hubs, homes, and final destinations~\cite{chiou2024firstlastmile, zuo2020firstlastmile}. These features make BSS valuable for improving transit accessibility and reducing mobility gaps in underserved neighborhoods~\cite{berke2024access, babagoli2019exploring, mohiuddin2023does}.

New York City (NYC), characterized by extensive yet unevenly distributed transit networks and notable socioeconomic diversity, represents an important study area to examine transit equity issues.
Since its launch in 2013, NYC’s Citi Bike system has expanded from an initial 329 stations concentrated in central Manhattan to over 2,316 stations citywide by 2024~\cite{nycdotStreetsPlan2024}.
Despite substantial growth, Citi Bike stations remain predominantly located in central and transit-rich neighborhoods, leaving outer areas underserved~\cite{javid2023equity}. 
Recent studies continue to highlight mobility disparities in neighborhoods such as the Bronx, eastern Queens, and southern Brooklyn, which are less accessible compared to wealthier borough centers~\cite{nycdotStreetsPlan2024, berke2024access}. 

Strategic expansion of BSS into underserved communities requires careful planning for new station locations by considering (1) accurately predicting demand at newly planned (``cold-start'') stations and (2) effectively integrating predicted bike demand with existing transit accessibility measures. 
Traditional demand prediction methods, such as gravity-based models~\cite{simini2021deep} and spatial interpolation~\cite{liu2022coldstart}, rely on historical ridership data, limiting their usability for cold-start stations due to the absence of the ridership data.
Recent deep-learning approaches~\cite{liang2023deep} address this issue by integrating various types of built environment features.  
However, these methods use a simple concatenation of multimodal data, which may fail to capture the complex interactions among urban features, potentially affecting prediction accuracy.
Moreover, existing accessibility metrics like Public Transit Accessibility Level (PTAL)~\cite{tfl2015connectivity, adhvaryu2021public} primarily measure proximity and transit frequency but overlook realistic bike-to-transit potential usage, thus failing to provide practical insights into equitable accessibility.

To overcome these limitations, we introduce \textbf{Transit for All (TFA)}, a spatial computing framework designed to guide equitable BSS expansions. TFA has three components: (1) predicting bike demand at cold-start stations, (2) assessing realistic transit accessibility, and (3) recommending bike station placement.

First, to improve demand prediction at cold-start stations, TFA utilizes multimodal region representation learning (RRL), that handles various geospatial data, including satellite imagery, point of interest (POIs), transit infrastructure, and neighborhood demographics. Unlike conventional methods that rely on the concatenation of raw features, RRL captures interactions among multiple datasets, addressing the complexity of urban dynamics between modalities.
Second, to enhance traditional accessibility assessment methods, we introduce a novel weighted Accessibility Level (wAL), which integrates predicted bike-sharing demand with the conventional PTAL metric. This integration captures realistic neighborhood-level differences in transit accessibility that traditional metrics may overlook.
Third, based on our assessment via wAL, TFA strategically recommends new bike station locations that enhance equity by considering underserved areas characterized by high predicted demand and low existing transit accessibility. 

We validate TFA's effectiveness through a detailed case study in NYC. 
First, our analysis confirms that the RRL approach provides more accurate predictions for cold-start bike station demand compared to the baseline method that uses concatenated raw features. 
Second, our proposed wAL metric identifies significant yet previously unrecognized disparities in transit accessibility. Neighborhoods with predominantly Black, Hispanic, or medium-to-low-income populations indicate substantially lower wAL scores than major or high-income neighborhoods, meaning they have poorer accessibility compared to more affluent or centrally located neighborhoods.
Finally, we quantitatively demonstrate that strategic station placement guided by TFA substantially reduce transit accessibility disparities across income and racial groups by using the Gini coefficient~\cite{dorfman1979formula}.

These findings demonstrate TFA's practicality as an equity-centric decision-support tool, helping planners with actionable insights to balance ridership objectives and equity improvements. While we conduct a case study in NYC, TFA is broadly applicable to other metropolitan areas with similar urban characteristics, directly advancing global efforts toward equitable and sustainable transportation aligned with SDG 11.2. We summarize our main contributions:
\begin{itemize}
    \item \textbf{C1} We introduce \textit{Transit for All}, a spatial computing framework that predicts bike-sharing demand at cold-start stations by leveraging multimodal region representation learning.
    \item \textbf{C2}: We propose a novel metric, the \textit{weighted Accessibility Level (wAL)}, which integrates traditional transit accessibility metrics with predicted bike-sharing demand. This approach enables urban planners to strategically identify optimal locations for equitable bike station placements.
    \item \textbf{C3}: Through a comprehensive spatial analysis in NYC, we uncover previously unrecognized equity gaps across socio-economic neighborhoods and demonstrate how strategic BSS expansion guided by TFA can effectively improve equitable transit accessibility. Our results provide actionable insights for city planners to enhance equitable urban mobility.
\end{itemize}

%% file: section/data.tex
\section{Study Area}

We select New York City (NYC) as our study area due to its disparities in transit infrastructure, especially in lower-income and minority neighborhoods~\cite{mta2025queensbus, baghestani2024equity}, and the potential for bike-sharing systems (BSS) to address these inequities~\cite{babagoli2019exploring, berke2024access}. Although the city supports an extensive transit network, many outer-borough and low-income areas are classified as ``transit deserts'', where transit service is sparse or unreliable~\cite{jin2024bikesharing}. 
These transit desert areas disproportionately impact minority communities and low-income residents, limiting access to employment and essential services~\cite{jomehpour2020transit_deserts}. 

Furthermore, bike-share infrastructure remains less accessible to lower-income communities~\cite{mahajan2024global}. For example, NYC is one of the cities that remains continuing spatial inequities with the station distribution dominating in affluent areas~\cite{berke2024access}. 
These findings highlight the persistent mismatch between transit supply and underserved demand.
Given that over 600,000 daily Citi Bike rides occur in NYC~\cite{gong2024deciphering}, BSS clearly plays a role in travel patterns. However, its uneven distribution limits its impact on transit-poor neighborhoods. As a result, strategically placing new stations in these underserved areas offers a promising avenue to fill transit gaps and enhance equitable mobility, making NYC a fitting setting for our study.

\subsection{Distribution of Cold-start Stations in NYC}

\begin{figure}[h]
    \centering
    \includegraphics[width=\linewidth]{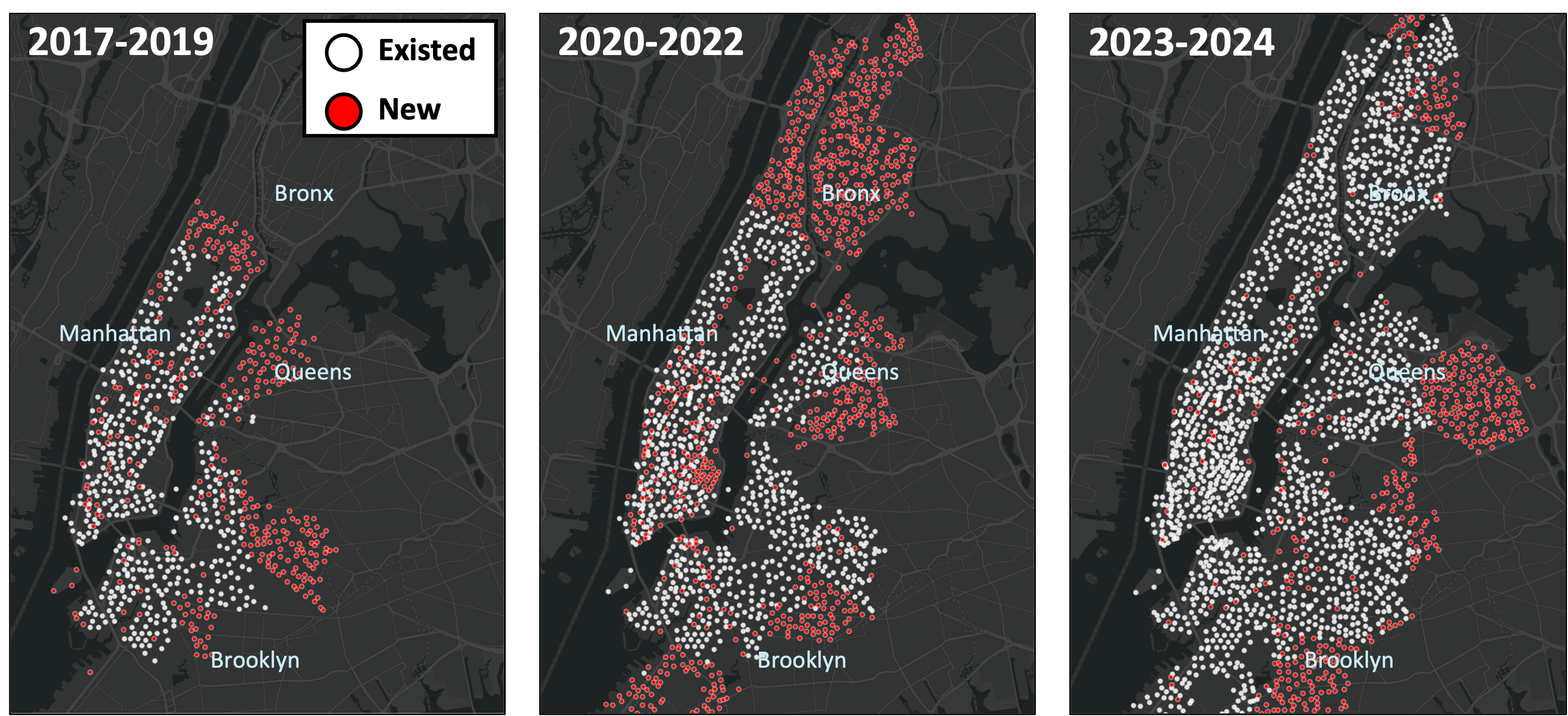}
    \caption{Spatial distribution of existing (white) and newly opened (red) BSS stations across NYC. Before 2017, 476 stations (white in 2017-2019) existed in Manhattan and Brooklyn. Between 2017 and 2019, 341 new stations (red) opened in Manhattan, Brooklyn, and Queens. From 2020 to 2022, 841 new stations (red) primarily expanded into upper Manhattan and the Bronx. Between 2023 and 2024, 494 new stations (red) further opened, mainly in Queens and southern Brooklyn.}
    \label{exp1:fig1}
\end{figure}

We define newly planned (``cold-start'') BSS stations, which have no previous records in a given month. 
\autoref{exp1:fig1} shows the spatial distribution of Citi Bike stations across three periods in NYC. 
In 2017, 476 stations (shown in white in the 2017-2019 image in Figure~\ref{exp1:fig1}) existed, predominantly in Manhattan and upper Brooklyn. Between 2017 and 2019, Citi Bike expanded, adding 341 new stations (shown in red), primarily expanding to Queens and Brooklyn, increasing accessibility across residential neighborhoods.

From 2020 to 2022, 841 new stations opened in upper Manhattan and the Bronx. This growing number of new stations aligns with a decision made by NYC’s Department of Transportation to enhance accessibility, especially near major hospitals, to support employees working at hospitals during the COVID-19 pandemic~\cite{nycDOT2020}. During 2023-2024, Citi Bike has further increased the BSS network by adding 494 stations. These new installations mainly cover previously undeveloped regions, including southern Brooklyn and eastern Queens. 
This recent expansion promotes transportation equity by improving first- and last-mile connectivity in neighborhoods that had limited access to BSS.

\begin{figure}[!h]
    \centering
    \includegraphics[width=\linewidth]{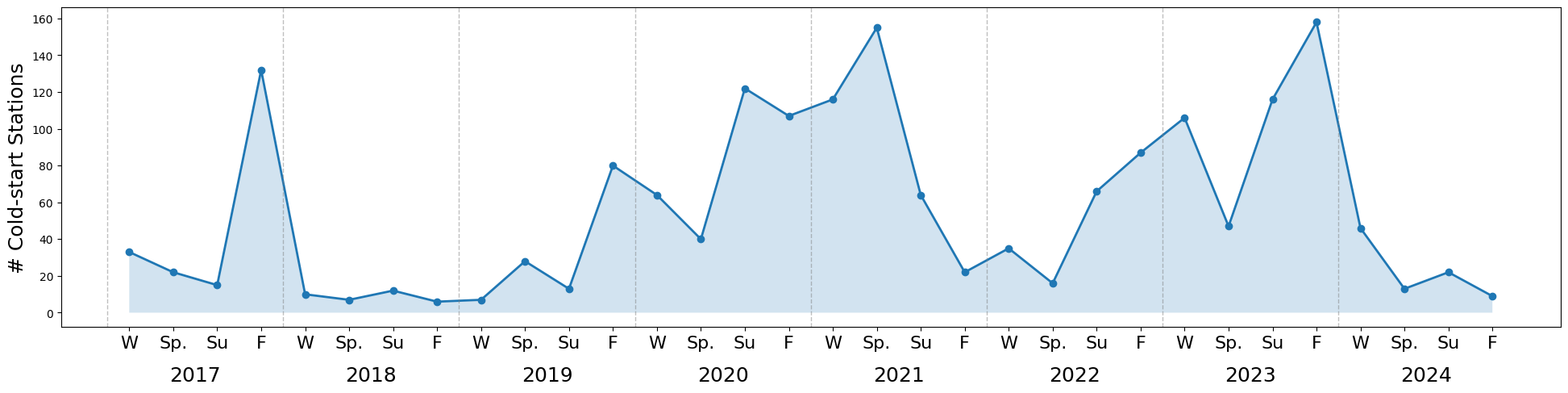}
    \caption{The number of newly opened (cold-start) bike stations by season from 2017 to 2024. Seasons are grouped as Winter (W: Dec–Feb), Spring (Sp: Mar–May), Summer (Su: Jun–Aug), and Fall (F: Sep–Nov).}
    \label{exp1:fig2}
\end{figure}

Further, we analyze seasonal trends in Citi Bike station openings. In Figure~\ref{exp1:fig2},
we observe distinct seasonal variations with notable peaks occurring during fall seasons (e.g., approximately 140 new stations in Fall 2020). Notably, 2020 exhibited high openings during summer with a total of around 340 new station installations. This surge aligns closely with city policies implemented to support increased bike usage among healthcare and essential workers during transit disruptions caused by COVID-19~\cite{nycDOT2020}. 

Overall, the spatial distribution and seasonal patterns of station openings illustrate how Citi Bike’s network has expanded from central Manhattan toward broader geographic coverage.


%% file: section/related_works.tex
\section{Literature Review}

\begin{figure*}[!h]
    \centering
    \includegraphics[width=\textwidth]{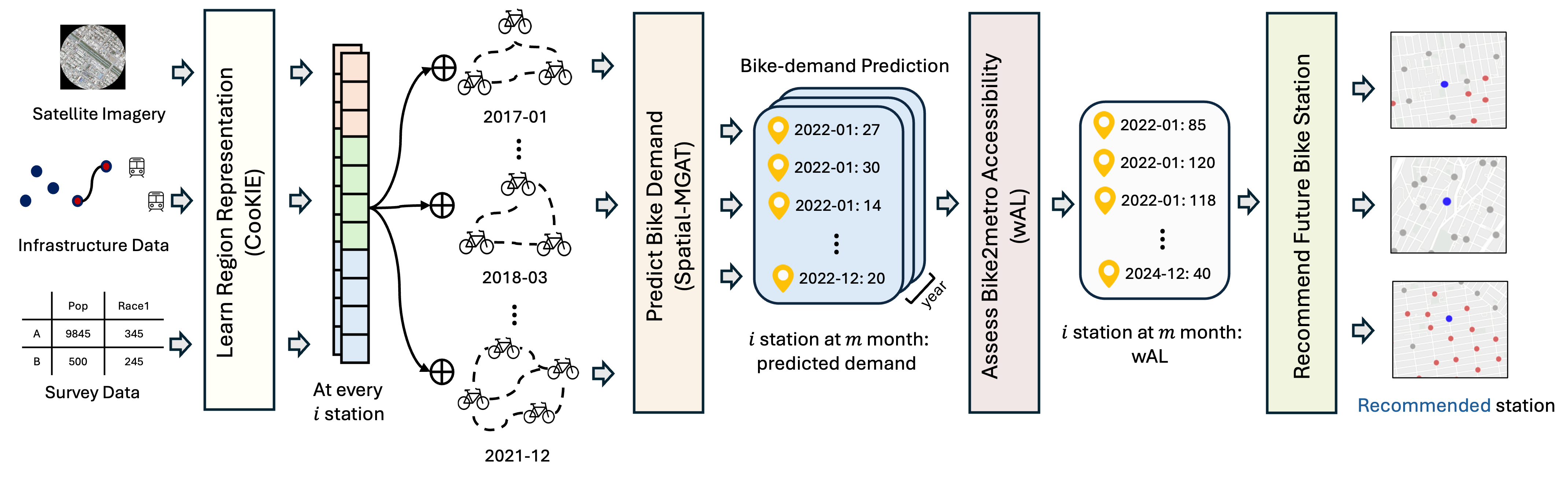}
    \caption{An overview of our Transit for All (TFA) framework. TFA consists of four components: (1) Learning region representaion via \modelnamex, (2) Predicting bike demand at each cold-start station using \predictorx, (3) Assessing bike to transit accessibility by computing wAL, and (4) Recommending future bike station placement.}
    \label{analysis:framework}
\end{figure*}

\subsection{Mobility Infrastructure Planning Frameworks}
Urban planners and policymakers increasingly use diverse transportation data to enhance equitable urban mobility. Existing frameworks commonly predict BSS demand using regression models informed by built environment characteristics, evaluate pedestrian accessibility based on walking distances, and suggest new station locations~\cite{chen2024locating, beairsto2022identifying, mix2022optimal}. While these studies consider equity in station placement, they focus on pedestrian-focused accessibility, neglecting an integration with existing transit networks with BSS, such as subways or buses, limiting multimodal connectivity.

Some recent works examine multimodal transportation integration. One study investigates equity improvements from integrating bike-sharing with metro services in Nanjing, China, quantifying accessibility gains, yet it does not provide recommendations for future station placement~\cite{cheng2024bike}. 
Another analyzes bike-to-rail transfers in Washington, D.C.’s bike-share and metro system, identifying factors influencing usage patterns. However, they neither evaluate equity impacts nor propose strategies for equitable expansion~\cite{tushar2024bikeshare}.

In contrast to these studies, our \textit{Transit for All} framework advances multimodal integration by (1) predicting demand at cold-start bike stations using multimodal representation learning, (2) introducing a novel weighted transit accessibility metric (wAL) that combines predicted bike demand and existing public transit accessibility, and (3) strategically recommending equitable BSS expansions designed to improve first- and last-mile transit connectivity in underserved communities.

\subsection{Transit Accessibility Measurements}
Transit accessibility measurements quantify how easily residents can reach public transportation from given locations. Traditional metrics such as the Accessibility Level (AL) evaluate accessibility based on proximity to transit infrastructure and scheduled transit frequencies~\cite{tfl2015connectivity}. 

Researchers adopt AL in cities beyond London (United Kingdom), including Lucknow (India)~\cite{shah2016public} and Hubli-Dharwad (India)~\cite{adhvaryu2021public}, supporting infrastructure planning in a variety of urban contexts. While AL identifies areas near transit stations, this approach does not incorporate actual or predicted travel demand. As a result, AL alone may not fully reflect the real-world transit needs of communities.

Recent approaches address this limitation by integrating observed transit demand into accessibility metrics. For instance, demand weighted methods combine ridership data with infrastructure availability, helping to identify mismatches between transit supply and actual usage~\cite{yen2023public}. However, these methods often neglect multimodal transportation connections, such as the connectivity between bike-sharing services and subway networks.

To bridge this gap, we introduce a weighted AL (wAL) metric that extends traditional AL by integrating factors such as bike travel time, waiting time, and transit frequency with predicted bike-sharing demand. By jointly considering infrastructure and expected usage, wAL enables planners to strategically place BSS stations and advance equitable transit accessibility across urban neighborhoods.

%% file: section/method.tex
\section{Transit for All}
We propose \textit{Transit for All (TFA)}, a spatial computing framework designed to guide equitable BSS expansions. In Figure~\ref{analysis:framework}, 
we show the overall TFA framework, which integrates three main components: (1) predicting demand at cold-start bike stations, (2) assessing transit accessibility at cold-start stations, and (3) recommending strategic bike station locations.

First, we apply region representation learning (RRL), utilizing multimodal data, such as satellite imagery, points-of-interest, demographics, and existing transit infrastructure, to generate region vector and predict bike demand with a supervised predictor.
Second, we introduce a novel accessible metric, the weighted Accessibility Level (wAL), which integrates predicted ridership with transit accessibility to better evaluate station locations.
Finally, we recommend future bike station placement locations based on wAL scores, enabling planners to address transit equity gaps in underserved communities.

\subsection{Predict Bike Demand at Cold-start Stations}\label{sec:frame1}
\subsubsection{Region Representation Learning}
RRL encodes geospatial data into compact embeddings representing geographic areas. We use the Cross-modal Knowledge Injected Embedding (\modelnamelogo)~\cite{namgung2025less} to integrate multimodal data, such as satellite imagery, urban infrastructure, and socio-demographics into unified region embeddings.
Existing RRL approaches often concatenate raw features directly, which may overlook the complementary relationships between modalities. For example, satellite imagery captures spatial patterns, socio-demographics describe community characteristics, and infrastructure reflects land functions. Direct concatenation may miss these complementary multimodal relationships.

To learn these interactions, \modelname uses two learning objectives: (1) \textbf{intra-view learning}, which learns region-specific features by comparing with other regions and (2) \textbf{pairwise inter-view learning}, which captures relationships across modalities.

\textbf{Intra-view learning} generates embeddings within each modality by using other regions for capturing unique region features. We apply contrastive learning (CL) using the InfoNCE loss~\cite{oord2018representation}:
\begin{equation}\label{eq:L_intra}
\mathcal{L}_{intra} = \sum_{i=1}^m -\log \frac{\exp(z_i \cdot z_i^{+} / \tau)}{\exp(z_i \cdot z_i^{+} / \tau) + \sum_{j=1}^{N} \exp(z_i \cdot z_{i,j}^{-} / \tau)},
\end{equation}
where $z_i$ is the representation of a region from an encoder at cold-start station $i$, $z_i^+$ is a positive augmented embedding of the same region, $z_{i,j}^-$ are embeddings from other regions, and $\tau$ is the temperature parameter in CL. Minimizing $\mathcal{L}_{intra}$ encourages representations of the same region to be similar and distinguishes them clearly from other regions.

\textbf{Pairwise inter-view learning} then leverages embeddings from intra-view learning to explicitly model multimodal interactions by learning shared and modality-specific information:
\begin{equation}
\mathcal{L}_{inter} = \mathcal{L}_{inter_u} + \mathcal{L}_{inter_s},
\end{equation}
where $\mathcal{L}_{inter_u}$ capture the uniqueness of each modality by minimizing redundancy, and $\mathcal{L}_{inter_s}$ learn shared information across modalities. Detailed derivations of these terms in~\cite{namgung2025less}.

\modelname jointly optimizes these objectives using a combined loss:
\begin{equation}
\mathcal{L} = \alpha \cdot \mathcal{L}_{intra} + \mathcal{L}_{inter},
\end{equation}
where $\alpha$ balances intra-view and inter-view learning.
Through joint optimization, \modelname produces a region representation that encodes geographic context and multimodal interactions. The region representation for a cold-start station $i$ is thus obtained by:
\begin{equation}
x_i^c = \text{\modelname}(x_i^1, x_i^2, x_i^3),
\end{equation} 
where \( x_i^1, x_i^2, x_i^3 \) denote multimodal input features from satellite imagery, urban infrastructure, and socio-demographics (see details in Section~\ref{imp}), and \( x_i^c \) denotes the final learned region embedding used for predicting bike demand at station \( i \).

\subsubsection{Predictor}
After extracting the learned region representation via \modelnamex, we feed it into our predictor, \predictorx~\cite{liang2023deep}.  
Bike demand at a newly planned station is often influenced by two types of existing stations: (1) nearby stations with similar travel patterns, and (2) stations that share similar built environment characteristics. 

To capture these dependencies, we denote by $m$ the month (time step) and construct two localized graphs for each target station 
$i$:
\begin{itemize}
    \item \( G^{p}_{i,m} \): a proximity-based graph that links the target station $i$ to its $k$ geographically nearest stations ($p$ denotes geographical proximity).
    \item \( G^{b}_{i,m} \): a built environment similarity graph  that connects stations $i$ to its $k$ most similar stations based on built-environment characteristics (e.g., POIs, transit infrastructure, and socio-economic attributes), where $b$ denotes built environment similarity. 
\end{itemize}
For each graph \( G^{*}_{i,m} (*\in \{ p, b \})\), we denote by \(\mathcal{N}^{(*)}_{i,m}\) the set of neighbor stations directly connected to the target stations $i$. 
From these graphs, we derive spatial-interaction vectors \( s^{(p)}_{i,m} \) and \( s^{(b)}_{i,m} \) by aggregating neighbor embeddings via learned attention:
\begin{equation}\label{eq:sim}
s^{(p)}_{i,m} = \sum_{j \in \mathcal{N}^{(p)}_{i,m}} \epsilon^{(p)}_{ij,m} \cdot h_{j,m},\quad\quad
s^{(b)}_{i,m} = \sum_{j \in \mathcal{N}^{(b)}_{i,m}} \epsilon^{(b)}_{ij,m} \cdot h_{j,m}.
\end{equation} 
Here,
\begin{itemize}
    \item \( h_{j,m} \) is the embedding of a neighbor station $j$ in month $m$, computed as 
    \begin{equation}
         h_{j,m} = \text{MLP}([x_j^c \parallel x_{j,m}^u]),
    \end{equation} 
    where \([x_j^c \parallel x_{j,m}^u]\) concatenates the learned region embedding $x_j^c$ from \modelname with the monthly urban context features $x_{j,m}^u$.
    \item \( \epsilon^{(*)}_{ij,m}  \) is the attention coefficient that measures the relative importance of neighbor $j$. Specifically,
    \begin{equation}
        \epsilon^{(*)}_{ij,m} = 
        \frac{\exp(\text{Attn}(W \cdot h_{i,m} \parallel W \cdot h_{j,m}))}{
        \sum\limits_{k \in \mathcal{N}^{(*)}_{i,m},\, k \neq i} 
        \exp(\text{Attn}(W \cdot h_{i,m} \parallel W \cdot h_{k,m}))
        },
    \end{equation}
    where \( W \) is a shared linear transformation parameter, and \(\text{Attn}(\cdot)\) is a two-layer feed-forward network computing attention scores.
\end{itemize} 



Finally, we concatenate each station’s \modelnamex's embedding \( x_i^c \), monthly urban context \( x_{i,m}^u \), temporal features \( t_{i,m} \), and spatial interaction vectors from the two graphs to predict the bike demand with a two-layer feed-forward neural network:
\begin{equation}\label{eq:mlp_chain}
\hat{y}_{i,m} = \mathrm{MLP}_2\left(\mathrm{MLP}_1\left(\mathbf{x}^{\text{cat}}_{i,m}\right)\right),
\end{equation}
where $\mathbf{x}^{\text{cat}}_{i,m}$ is a concatenation of $x_i^c$, $x_{i,m}^u$, $t_{i,m}$, $s^{(p)}_{i,m}$ and $s^{(b)}_{i,m}$.

\subsection{Assess Public Transit Accessibility}
To better reflect realistic transit accessibility and guide equitable infrastructure planning, we propose the weighted Accessibility Level (wAL) by integrating predicted bike demand from \predictorx. 
Traditional metrics like the Accessibility Level (AL), developed by Transport for London~\cite{tfl2015connectivity}, primarily measure accessibility based on transit proximity and service frequency.

\textbf{Accessibility Level (AL):}  
AL quantifies accessibility by aggregating subway access frequency via the Equivalent Doorstep Frequency (EDF), which integrates bike travel time and subway waiting periods. EDF is calculated by:
\begin{equation}
EDF_{ij} = \frac{30}{t_{bike_{ij}} + 0.5 \times \frac{T}{N_j} + 0.75}, \label{eq:edf_combined}
\end{equation}
where:
\begin{itemize}
    \item $t_{bike_{ij}}$ is bike travel time (minutes) from cold-start station $i$ to subway entrance $j$. $t_{bike_{ij}}$ is obtained by converting the shortest path distance (using Dijkstra's algorithm~\cite{dijkstra1959note}) to minutes, assuming an average cycling speed of approximately 280 meters per minute (16.8 km/h)~\cite{yen2023public}.
    \item $\frac{T}{N_j}$ refers to average interval between trains at train entrance $j$. $T$ is the observation period (e.g., 7:30–9:30 AM, peak commuting hours~\cite{tfl2015connectivity}) and $N_j$ is the number of scheduled trains at subway station $j$ within period $T$.
    \item $0.5 \times \frac{T}{N_j}$ estimates average scheduled waiting time.
    \item $0.75$ minutes is a buffer accounting for service variability~\cite{tfl2015connectivity}.
\end{itemize}
We then aggregate EDF values, selecting the subway entrance with the highest accessibility and weighting secondary entrances with half importance:
\begin{equation}
AL_i = \max_j(EDF_{ij}) + 0.5 \times \sum_{j \neq j_{max}} EDF_{ij}.
\end{equation}

However, AL alone does not account for potential bike demand at cold-start stations, which is obtained from Section~\ref{sec:frame1}. For example, two stations may have the same AL, but if one has much lower bike demand, its accessibility will translate into far less actual use.

\textbf{weighted Accessibility Level (wAL):} Our proposed metric enhances traditional AL by incorporating predicted bike demand ($\hat{y}_{i,m}$) from \predictorx. Traditional AL solely considers accessibility based on proximity and service frequency without considering the potential utilization by local residents. For example, consider two neighborhoods with identical AL scores. The neighborhood with higher predicted bike usage would likely experience greater practical benefits from the bike-sharing station compared to the neighborhood with lower predicted usage, as a larger number of residents are expected to use bikes to connect to transit services. Thus, wAL captures not only how accessible a station is theoretically (transit proximity and frequency) but also how beneficial or useful it will realistically be to the community (predicted demand):

\begin{equation}
wAL_i = AL_i \times \hat{y}_{i,m},
\end{equation}
where $\hat{y}_{i,m}$ is predicted bike demand at cold-start station $i$, and $AL_i$ is the AL score at $i$.

\subsection{Recommend New Bike Station Placement}\label{method:future}
To address gaps in equitable transit accessibility, we recommend locations for new bike-sharing stations in NYC based on wAL. An official Citi Bike guidelines for bike station placement~\cite{NYCDOT2013BikeShare} focus on both geographic information and site-specific criteria. For example, they include minimum sidewalk widths (16 ft), curb-lane widths (8 ft), spacing of approximately 305 meters (1,000 ft) between stations, a 15-ft clearance from subway entrances, restrictions around fire hydrants, and bus stops, and several on-site observations, such as avoiding building entrances and construction area. However, these guidelines do not incorporate demand prediction~\cite{berke2024access, mahajan2024global, javid2023bikeshare_equity}, which is critical for ensuring that new stations address local community transit needs and maximize their actual utilization.

To integrate equitable transit accessibility into future station placement decisions, we propose a strategic recommendation method guided by:
\begin{itemize}
    \item \textbf{Selecting candidate neighborhood:} We first find candidate neighborhoods in census tracts characterized by lower median incomes and higher proportions of minority populations~\cite{baghestani2024equity, jomehpour2020transit_deserts}.
    \item \textbf{Walkable road accessibility:} We leverage OpenStreetMap (OSM) road network data to ensure stations are reachable by walk, such as \textit{residential streets, living streets, tertiary, and secondary roads}, for enhancing accessibility.
    \item \textbf{Consistent with spacing guidelines:} Consistent with Citi Bike guidelines~\cite{NYCDOT2013BikeShare}, we maintain a minimum spacing of approximately 305 meters (1,000 feet) between stations to avoid redundancy and ensure efficient network coverage.
    \item \textbf{Considering expected bike usage and transit accessibility:} 
    We select candidate station locations with wAL. Stations with higher wAL scores indicate areas where improved bike-sharing access would significantly enhance equitable transit accessibility and community-level mobility.
\end{itemize}

%% file: section/experiment.tex
\section{Case Study: New York City}
\subsection{New Station Demand Prediction Data}
We use open-source geospatial data to characterize the built environment within a 500-meter buffer around each cold-start station.

\subsubsection{\modelname + \predictor input data:} \label{input_data_sec}
Following~\cite{liang2023deep}, we select these neighborhood features: 
\begin{itemize}
    \item \textbf{Satellite imagery}: We sample high-resolution (0.6-meter) aerial imagery from the 2021 National Agriculture Imagery Program (NAIP)~\cite{naip2021} within a 500-meter buffer around each cold-start station.

    \item \textbf{Road network (7 features)}: We calculate the total lengths of seven road classes, motorway, trunk, primary, secondary, tertiary, unclassified, and residential, within a 500-meter buffer using OpenStreetMap (OSM)~\cite{geofabrik2025}.

    \item \textbf{Point-of-Interest (POI) (10 features)}: We count POIs within 500 meters across ten categories (residential, educational, cultural, recreational, commercial, religious, transportation, government, health, social) using NYC Open Data~\cite{nycCommonPlace2025, nycPOIMetadata2025}.
    
    \item \textbf{Subway accessibility (2 features)}: We extract (1) the number of subway stations within 500 meters and (2) the distance to the nearest subway station from NYC metro data~\cite{mtasubway2025}.
    
    \item \textbf{Socio-demographic attributes (10 features)}: We integrate population density and ethnicity distribution data from the 2020 U.S. Census Survey~\cite{uscensus2025}.
    
    \item \textbf{Monthly dynamic features}: For \predictorx, we compute monthly-updated features, including bike lane lengths from OSM~\cite{geofabrik2025}, number of existing bike stations within 500-meters, 1000-meters, and between 1000- to 5000-meters using OSM~\cite{geofabrik2025}, average travel distance from nearby existing BSS stations, and station age. 
    
\end{itemize}

\subsubsection{wAL input data}
To calculate wAL scores for cold-start stations, we combine predicted bike demand and AL score. To compute AL, we utilize subway entrances, pedestrian road networks, and transit schedules data. Specifically, we collect:
\begin{itemize}
    \item \textbf{Subway entrances}: We aggregate geographic locations of subway entrances from NYC Open Data~\cite{mtasubway2025} within a 500-meter buffer around each cold-start station.
    
    \item \textbf{Road network}: We construct a detailed pedestrian network using road data from OpenStreetMap (OSM) via Geofabrik~\cite{geofabrik2025} to accurately model pedestrian accessibility.
    
    \item \textbf{MTA subway schedule (GTFS Data)}: We leverage subway schedules, service intervals, and frequencies from the MTA’s General Transit Feed Specification (GTFS) dataset~\cite{mtagtfs2025}.

\end{itemize}


\subsection{Training Details on Bike Demand Prediction}\label{imp}
We generate region embeddings using \modelname for 2,316 bike stations based on a 500-meter buffer around each station. Our analysis includes only BSS stations, which is active as of December 2024. We train \predictor using BSS ridership data from January 2017 to December 2021, and test \predictor on data from January 2022 to December 2024.
For detailed model training, we train \modelname for 1000 epochs with a 0.0001 learning rate. The hidden dimensions are 512. 
For \predictorx, we follow the model settings in \cite{liang2023deep}, using 200 training epochs, the Adam optimizer with a learning rate of 0.002, and the number of neighbors set to $k=5$.



%% file: section/result.tex
\section{Results \& Discussion}
We first compare bike demand prediction at cold-start station using \modelname + \predictor versus raw features + \predictorx. Next, we visualize wAL distributions across NYC. We then perform zone-level analyses with census data across multiple scenarios and functional areas. Finally, we recommend future locations for potential BSS stations.

\subsection{Demand Prediction at Cold-start Stations}
We design an experiment that feeds multimodal representation learned by \modelname and raw feature concatenation to \predictor as inputs. We evaluate BSS demand prediction at cold-start stations during the test period (See Section~\ref{imp}) using root mean squared error (RMSE), mean average error (MAE), and R-squared ($R^2$)~\cite{liang2023deep} in Table~\ref{tab1:result}.

Using the learned representation from \modelname yields substantial improvements: RMSE decreases by 7.40\%, MAE by 11.74\%, and $R^2$ improves by 6.55\%. These results indicate that \modelname effectively captures complex multimodal dependencies among region characteristics, providing spatial context compared to the simple raw feature concatenation method.

\begin{table}[htbp]
\centering
\caption{Performance comparison between learned representations from \modelname and raw feature concatenation as inputs to \predictorx. $\downarrow$ means the lower the better, while $\uparrow$ means the higher the better.}
\begin{adjustbox}{width=0.85\columnwidth,center}
  \begin{tabular}{l | c | c | c }  \toprule
  Methods & RMSE$\downarrow$ & MAE$\downarrow$ &$R^2$ $\uparrow$ \\ \midrule
  \modelname + \predictor & 27.975 & 13.554 & 0.667 \\ 
  Raw + \predictor & 30.210 & 15.357 & 0.626 \\ \midrule
  \textbf{Improvement \%} & \textbf{7.398} & \textbf{11.741} &  \textbf{6.550}\\
   \bottomrule
  \end{tabular}
  \label{tab1:result}
\end{adjustbox}
\end{table}

\begin{figure}[h]
    \centering
    \includegraphics[width=.95\linewidth]{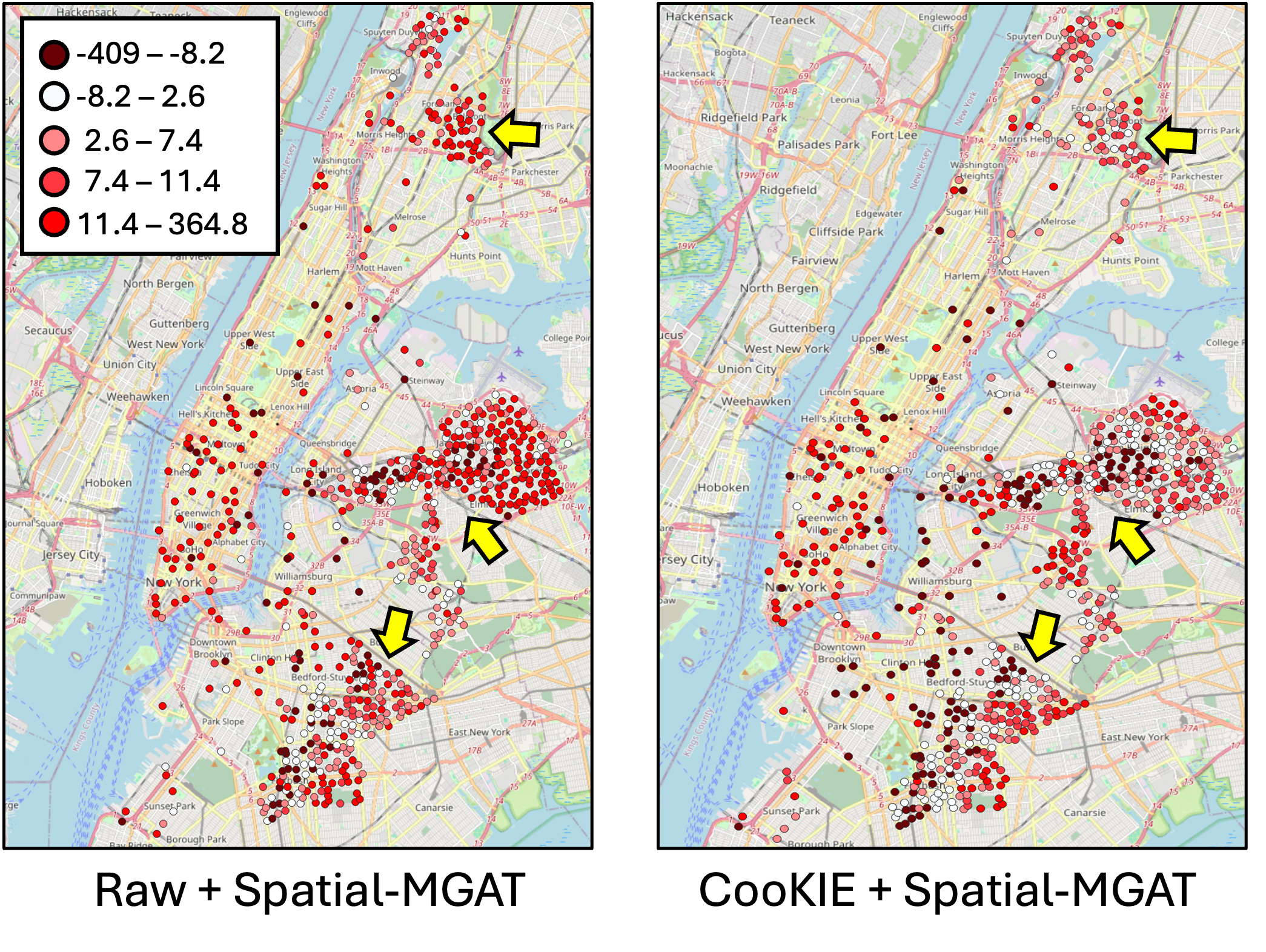}
    \caption{Spatial distribution of prediction errors for Raw + \predictor and \modelname + \predictor (Ours). Larger errors appear in dense regions (shown with arrows) when using raw features, whereas \modelname reduces these errors and yields more consistent predictions across the city.}
    \label{analysis:error_viz}
\end{figure}

Figure~\ref{analysis:error_viz} shows the prediction errors across NYC. With raw feature concatenation, large errors cluster in high-demand areas such as the Bronx, Queens, and Brooklyn (yellow arrows), where bike demand is difficult to capture from simple features. In contrast, \modelname reduces these errors, producing more stable predictions across diverse neighborhoods.

\subsection{Spatial Distribution of wAL}
\begin{figure}[h]
    \centering
    \includegraphics[width=.85\linewidth]{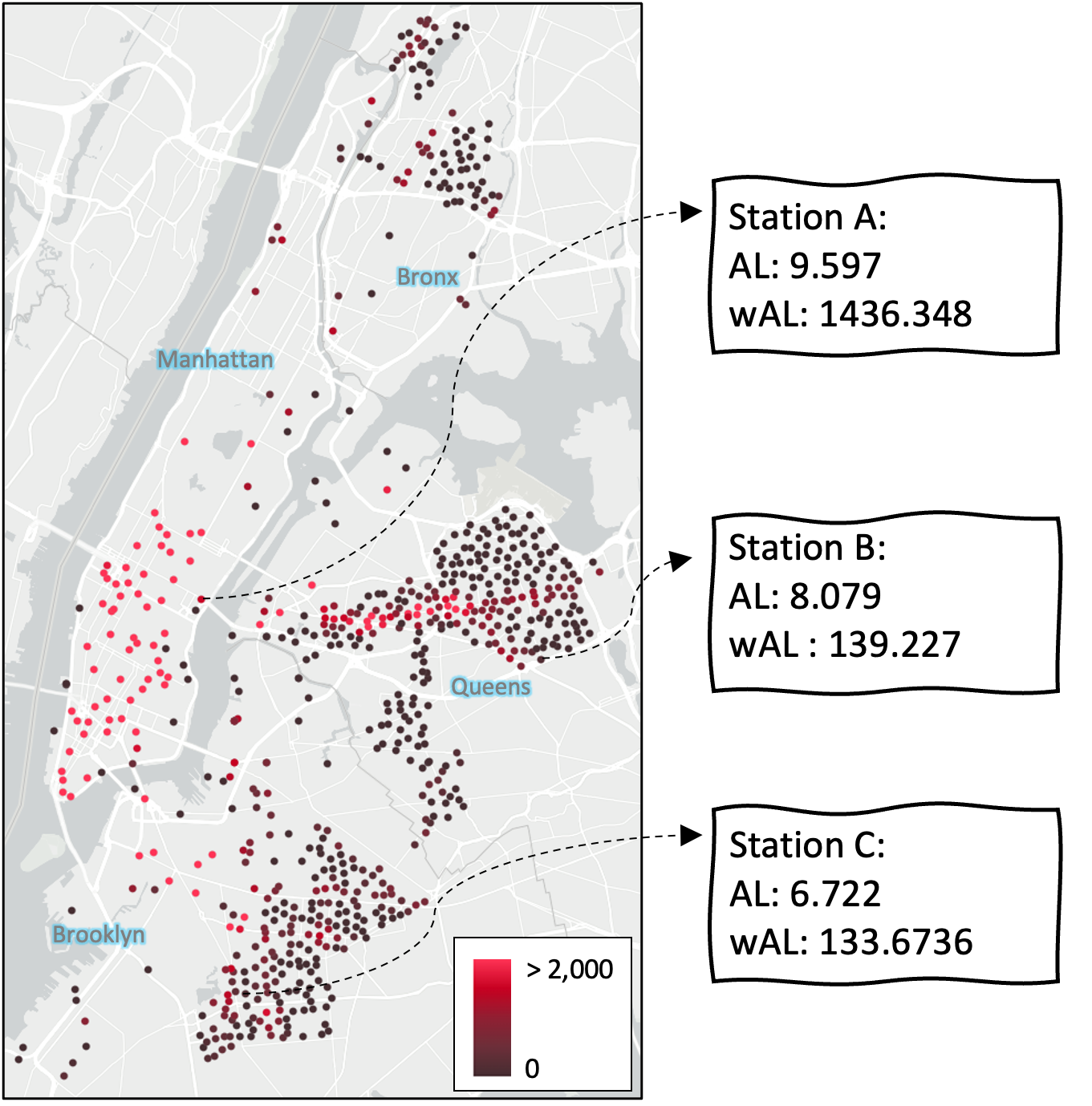}
    \caption{Spatial distribution of wAL for 699 new BSS stations during the test period (2022–2024). Red indicates higher wAL scores, identifying stations with both strong transit access and predicted ridership.}
    \label{analysis:fig1:wAL}
\end{figure}

We present the spatial distribution of wAL for newly planned BSS stations from January 2022 to December 2024 (testing period) in Figure~\ref{analysis:fig1:wAL}. Figure~\ref{analysis:fig1:wAL} shows Manhattan as having consistently high wAL scores, resulting from strong public transit infrastructure combined with high predicted ridership. In Queens, stations along major roads, particularly those aligned with subway lines, also exhibit high wAL values (shown in red), emphasizing the high transport accessibility with the demand. 
In contrast, Brooklyn shows more variability with high wAL scores concentrated near major transit hubs but notably lower scores in neighborhood areas. The Bronx shows similarly mixed accessibility patterns with moderate wAL values primarily clustered around central roads and transit corridors.

To clearly illustrate the advantage of using wAL over AL alone, we closely examine three representative stations: Station A (Manhattan), Station B (Queens), and Station C (Brooklyn). Although their raw AL scores are similar (differing by at most 2.8 points), their wAL scores vary significantly. Specifically, Station A achieves a wAL of 1,436.3, substantially higher than Stations B (139.2) and C (133.7). This substantial difference indicates that Station A, which is located in a densely populated area with high predicted ridership, benefits from integrating ridership predictions into accessibility metrics. Thus, incorporating predicted ridership into wAL effectively captures realistic differences in transit accessibility and provides guidance for equitable and impactful station placements.

\subsection{Zone-level Analysis across Scenarios}
We conduct an equity analysis~\cite{jiang2024large} of transit accessibility from cold-start stations using demographic and economic data from the 2020 U.S. Census and ACS~\cite{acs2023}. We focus on two variables: \textit{major ethnicity}, defined by the predominant ethnic group (White, Black, Asian, or Hispanic), and \textit{median household income}, categorized by four quantiles (high, med-high, med-low, low). We summarize variable definitions and groupings in Table~\ref{tab1:zonal_group}.

\begin{table}[htbp]
\centering
\caption{Demographic and Economic variables used for equity analysis across network scenarios.}
\begin{adjustbox}{width=0.95\columnwidth,center}
\begin{tabular}{l | l | l }  
\toprule
\textbf{Analysis Variable} & \textbf{Definition} & \textbf{Grouping} \\ 
\midrule
Major Ethnicity & Predominant ethnic group & White, Black,\\
& by population & Asian, Hispanic \\[6pt] \midrule

Median Income & Median household income & High, Med-high, \\
&  &Med-low, Low \\[6pt] 

\bottomrule
\end{tabular}
\label{tab1:zonal_group}
\end{adjustbox}
\end{table}




\subsubsection{Zone-level Analysis: wAL per Major Ethnicity}
\begin{figure*}[h]
    \centering
    \includegraphics[width=.95\textwidth]{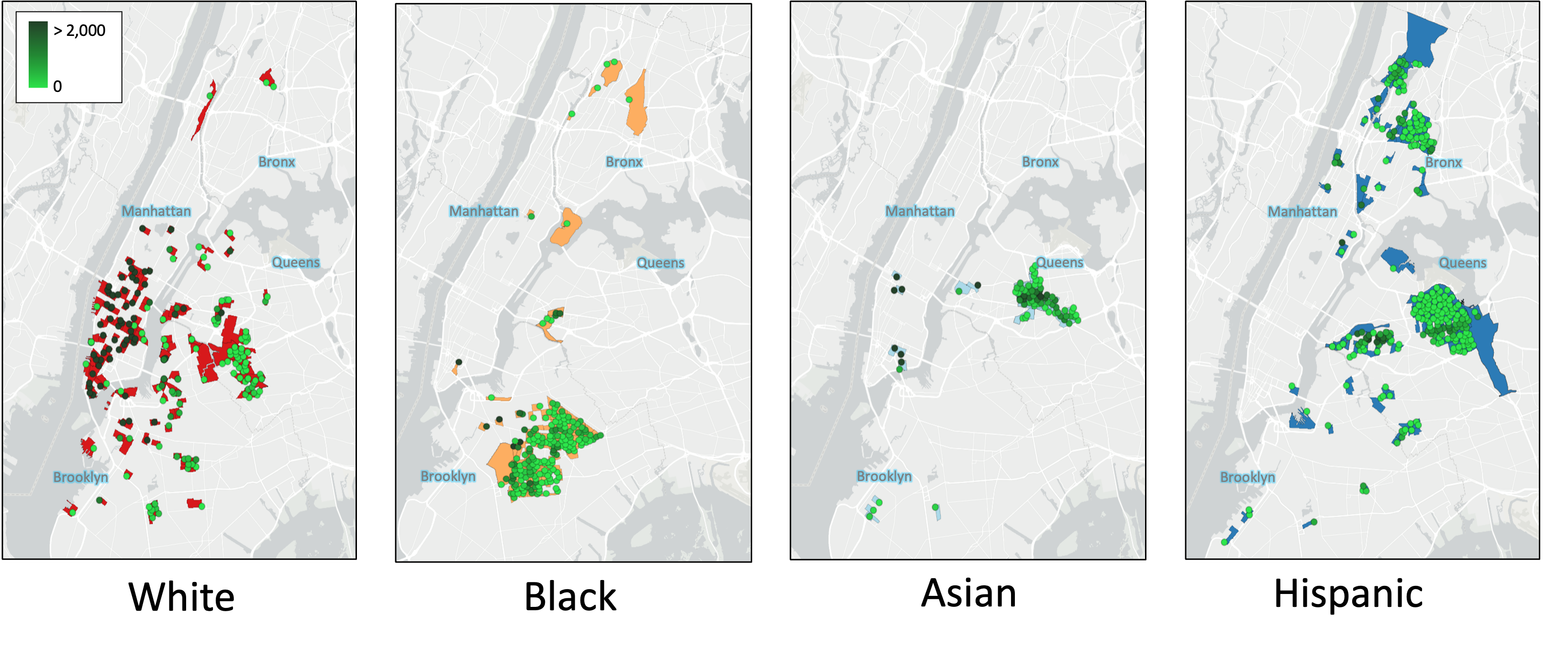}
    \caption{Spatial distribution of wAL for cold-start stations across major ethnic groups (White, Black, Asian, and Hispanic). Darker green indicates higher wAL scores, reflecting better public transportation accessibility. The map highlights distinct spatial accessibility patterns across ethnic groups.}
    \label{analysis:race}
\end{figure*}

\begin{figure*}[h]
    \centering
\includegraphics[width=.95\textwidth]{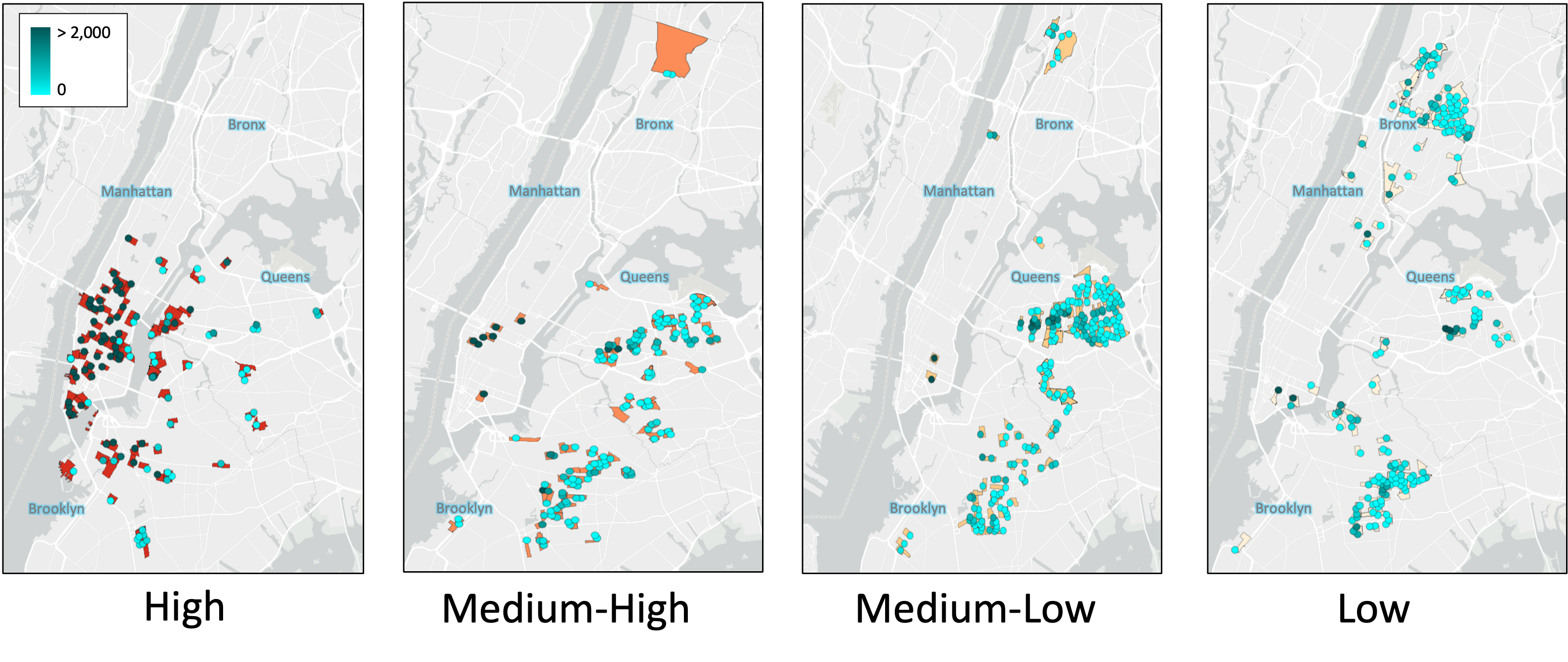}
    \caption{Spatial distribution of wAL for cold-start stations categorized by median income (high, medium-high, medium-low, low). Darker blue indicates higher accessibility scores.}
    \label{analysis:income}
\end{figure*}

We visualize the spatial distribution of wAL across four ethnic groups, White, Black, Asian, and Hispanic, in Figure~\ref{analysis:race}. We observe that the spatial distribution of wAL distinctly varies across these ethnicity groups. 
The White population predominantly resides in central Manhattan and along the boundary between Brooklyn and Queens. These areas benefit from extensive subway infrastructure, resulting in notably high accessibility with a mean wAL of 7,262.4. In contrast, the Black population is largely concentrated in southern Brooklyn, a region with fewer subway lines and limited public transit coverage, resulting in significantly lower accessibility with a mean wAL of 459.1. Asian neighborhoods cluster prominently in Queens, especially around major transit hubs such as Flushing and Jackson Heights neighborhood. These areas benefit from multiple subway lines and bus services, which yields relatively high accessibility reflected by a mean wAL of 2,138.3. 
Hispanic neighborhoods, predominantly living in northern Brooklyn and eastern Queens, typically lie further from major transit routes, resulting in the lowest mean wAL among the four ethnic groups at 244.0. 
Nonetheless, some central Hispanic neighborhoods in Queens exhibit better transit accessibility than those at the borough’s edges or the Bronx, indicating localized variations in service distribution.

Considerable variability within ethnic groups further highlights transit accessibility disparities: the standard deviation of wAL is notably high, such as 16,346.2 for White neighborhoods and 8,572.1 for Asian neighborhoods. These variations reflect disparities from not solely by ethnicity but also by neighborhood-specific urban planning and infrastructure.
Analyzing wAL at the borough level further clarifies these patterns. Black neighborhoods shows consistently low accessibility, with mean wAL scores of 324.4 in Brooklyn and 615.2 in Queens. This large difference in wAL by boroughs indicates significant transit gaps and the need for improvements to enhance equity. On the other hand, Asian neighborhoods in Queens show higher accessibility, with a mean wAL score of 652.1, benefiting from proximity to robust transit infrastructure. Hispanic neighborhoods show lower borough-level accessibility, with mean wAL scores of 183.4 in Brooklyn and 235.3 in Queens, highlighting transit limitations due to their distant residential locations relative to major transit locations.

\subsubsection{Zone-level Analysis: wAL per Median Income}
In Figure~\ref{analysis:income}, we analyze the spatial distribution of wAL by dividing regions into four median-income groups based on quantiles: high, medium-high, medium-low, and low. The resulting spatial patterns distinctly reflect the economic dynamics across NYC.

Stations in high median-income neighborhoods demonstrate the highest overall accessibility, with an average wAL of 9,503.7. These neighborhoods, predominantly located in central Manhattan, benefit significantly from extensive subway networks, frequent transit services, and robust infrastructure. High-income areas outside Manhattan, such as those in Brooklyn and Queens, still show relatively high accessibility but with notably lower mean wAL scores than in Manhattan: 1,130.9 in Brooklyn and 740.2 in Queens.

Neighborhoods with medium-high income have an average wAL of 2,180.5 citywide, though substantial variations exist among boroughs. Medium-high income neighborhoods in Manhattan have exceptionally high accessibility, averaging 38,264.0, due to proximity to major transit hubs and dense subway coverage. In contrast, neighborhoods in the same income category in Brooklyn and Queens represent considerably lower accessibility than in Manhattan, averaging 337.1 and 246.0, respectively. Such disparities within the same income group highlight significant differences in transit accessibility across boroughs.

Neighborhoods within the medium-low income category generally experience lower and more varied transit accessibility than high or medium-high categories, averaging a citywide wAL of 337.7. Within the medium-low income group, neighborhoods in Manhattan have comparatively higher accessibility than other boroughs, averaging 2,668.4, benefiting from nearby extensive transit infrastructure. However, medium-low income neighborhoods in the Bronx, Brooklyn, and Queens exhibit notably lower accessibility, averaging 172.6, 277.2, and 308.3, respectively. This large difference among boroughs underscores infrastructure disparities that significantly impact transit accessibility.

Finally, neighborhoods with low-income groups consistently experience the lowest accessibility citywide, averaging a wAL of 425.6. Borough-level analysis reveals particularly low accessibility scores for low-income neighborhoods in the Bronx at 205.7, Brooklyn at 260.9, and Queens at 392.4. In contrast, Manhattan’s low-income neighborhoods exhibit notably higher accessibility than those in three boroughs, averaging 2,221.8. These borough-level differences emphasize severe inequities in transit accessibility, disproportionately affecting economically vulnerable communities, particularly those in the outer boroughs compared to Manhattan.






\subsection{Zone-level Analysis across Functional Areas}
\begin{figure}[h]
    \centering
    \includegraphics[width=\linewidth]{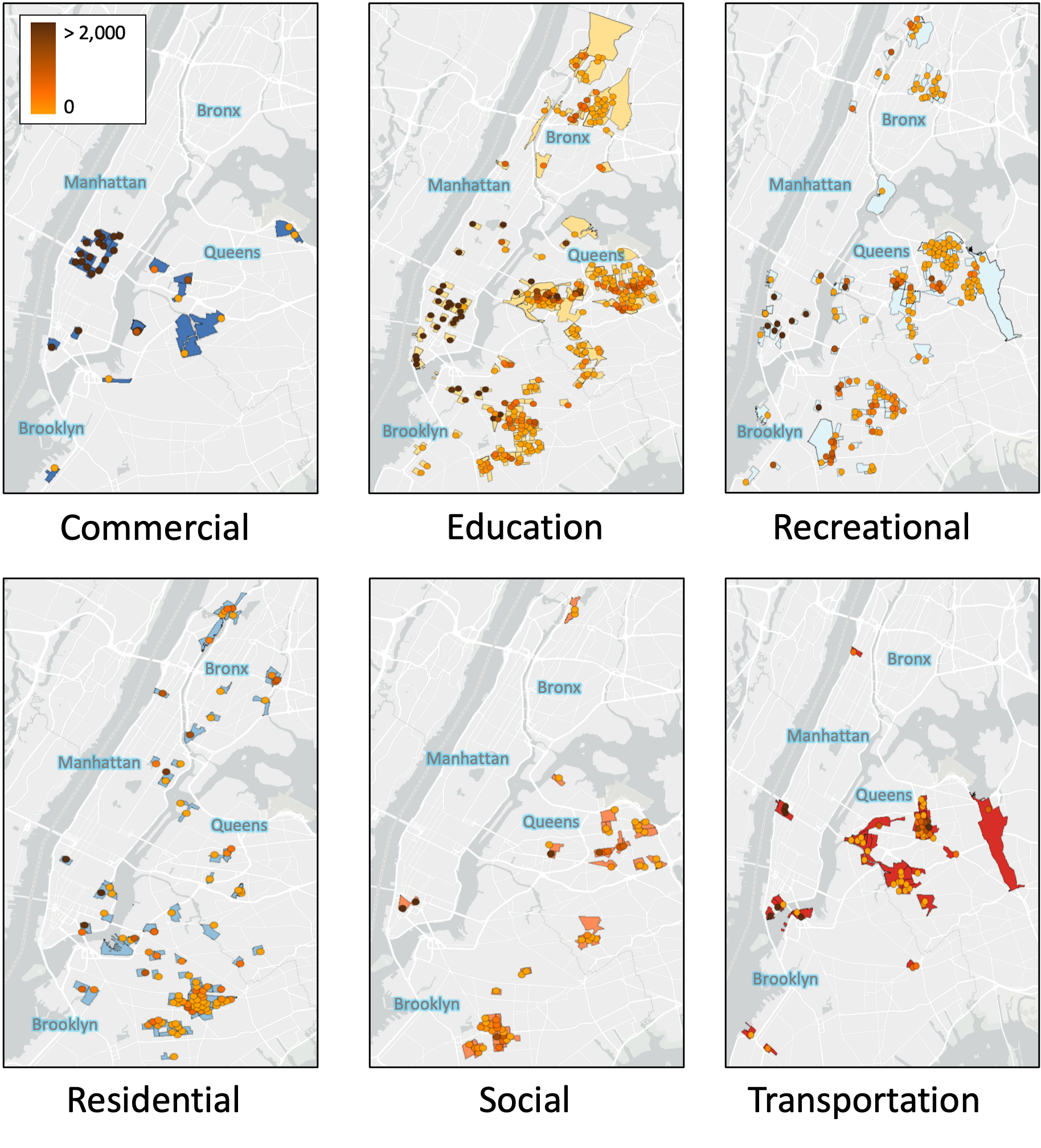}
    \caption{Spatial distribution of wAL by dominant Points-of-Interest (POIs): Commercial, Education, Recreational, Residential, Social, and Transportation. A darker orange color indicates higher wAL scores.}
    \label{analysis:poi}
\end{figure}

We analyze the spatial distribution of wAL across various POI categories: Commercial, Education, Recreational, Residential, Social, and Transportation zones, shown in Figure~\ref{analysis:poi}. The spatial patterns of wAL distinctly align with both demographic (Section~\ref{analysis:race}) and economic patterns (Section~\ref{analysis:income}) in NYC.

Commercial zones consistently exhibit high wAL, notably in Manhattan’s central business districts with an average wAL of 37,303.4. These areas benefit substantially from dense subway networks and frequent transit services. Brooklyn's commercial zones also have relatively high accessibility, averaging 873.0, whereas Queens shows significantly lower scores, averaging 305.5.
Transportation-related POIs generally display high wAL due to their connectivity and proximity to major transit hubs, such as transit terminals and ferry landings. 

Manhattan’s transportation zones indicate exceptionally high wAL, averaging 4,660.9. However, Queens and Brooklyn demonstrate markedly lower accessibility, averaging 409.5 and 579.0, respectively, reflecting variations in transit service frequency and density.
Residential zones typically exhibit lower overall wAL with borough-specific variations clearly visible. Residential areas in Manhattan show relatively high average accessibility scores at 1,790.4, benefiting from proximity to extensive transit infrastructure. In contrast, residential zones in Brooklyn, Queens, and the Bronx demonstrate significantly lower accessibility scores than Manhattan, averaging 195.3, 113.3, and 462.1, respectively. Particularly, low scores in southern Brooklyn and the Bronx align closely with areas predominantly by lower-income and minority communities.

Educational zones present varied transit accessibility, heavily influenced by proximity to major transit. Manhattan educational institutions, including universities and large public schools, show extremely high average accessibility at 19,004.0. Conversely, educational facilities in Brooklyn (577.7), Queens (426.7), and especially the Bronx (211.8) display notably lower wAL scores, reflecting disparities in educational accessibility linked to transit infrastructure. 

Recreational POIs similarly display significant variability. Central Manhattan recreational facilities, benefiting from proximity to transit-rich landmarks and attractions, average 5,592.7 in accessibility. However, recreational zones in outer boroughs such as the Bronx (89.6), Queens (204.2), and Brooklyn (426.8) indicate substantially lower scores, particularly for remote locations such as beaches or golf courses.
Social service zones, including daycare centers, nursing homes, and shelters, show notably low wAL across boroughs. Manhattan's social service facilities show high wAL averaging 23,274.0 due to central locations within transit-rich neighborhoods. Meanwhile, social service facilities in Queens (273.1) and Brooklyn (284.7) reflect lower accessibility, highlighting critical transit inequities impacting vulnerable populations.

\subsection{Future BSS Station Recommendations}
We quantify equity improvements in wAL for future bike stations using the Gini index~\cite{jiang2024large, allanson2014income}. The Gini index captures disparities in accessibility distribution, where a lower value indicates more equitable access across demographic (income) groups. The Gini index ($G$) measures relative inequality among demographic groups by comparing group means to overall weighted accessibility.
Specifically, the $G$ for a set of $M$ demographic groups is defined as:
\begin{align}
    G &= \frac{ \sum_{i=1}^M \sum_{j=1}^M w_i w_j |m_i - m_j| }{ 2W^2 \mu },
\end{align}
where $w_i$ and $w_j$ represent the number of stations in demographic groups $i$ and $j$ (e.g., income-based categories like high, low), respectively, $m_i$ and $m_j$ are the mean wAL score within groups $i$ and $j$, $W$ is the total number of stations across all demographic groups, and $\mu$ is the population-weighted overall mean accessibility.

\begin{figure}[h]
\centering
\includegraphics[width=.9\linewidth]{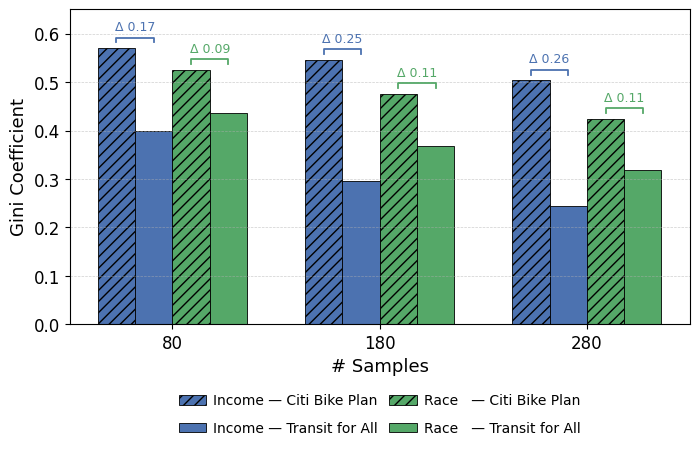}
\caption{Comparison of Gini coefficients measuring equity in wAL by income and race, Citi Bike Plan vs. Transit for All, incrementally adding recommended bike stations. Lower Gini coefficients indicate better equity.}
\label{analysis:rec}
\end{figure}

Using our \textbf{Transit for All (TFA)} framework, we recommend strategic locations for 280 future bike stations. Figure~\ref{analysis:rec} presents equity improvements by incrementally introducing new recommended stations (e.g., 80, 180, 280) to the existing 699 cold-start stations from the 2022–2024 testing period. We compare our approach with NYC’s Citi Bike installation plan~\cite{NYCDOT2013BikeShare} by excluding on-site observations. This baseline follows real-world scenarios by reflecting existing BSS expansion.

Based on the nearest distance to 699 existing stations, we select 80, 180, and 280 recommended bike stations to Citi Bike Plan and Transit for All (ours).
Adding 80 recommended stations reduces the income-based Gini coefficient significantly, by 0.17 (from 0.57 to 0.40), and improves the race-based Gini by 0.09 (from 0.52 to 0.44). 
Further, by continuously adding recommended stations (180 and 280 stations) to the existing stations, we observe that lowering disparities by 0.25 and 0.26 for income and consistently reducing racial disparities by 0.11. This consistent improvement in equity demonstrate that strategically selecting station sites using the TFA framework significantly enhances transit equity, systematically decreasing accessibility gaps among income and racial groups.

\begin{figure}[h]
    \centering
\includegraphics[width=.9\linewidth]{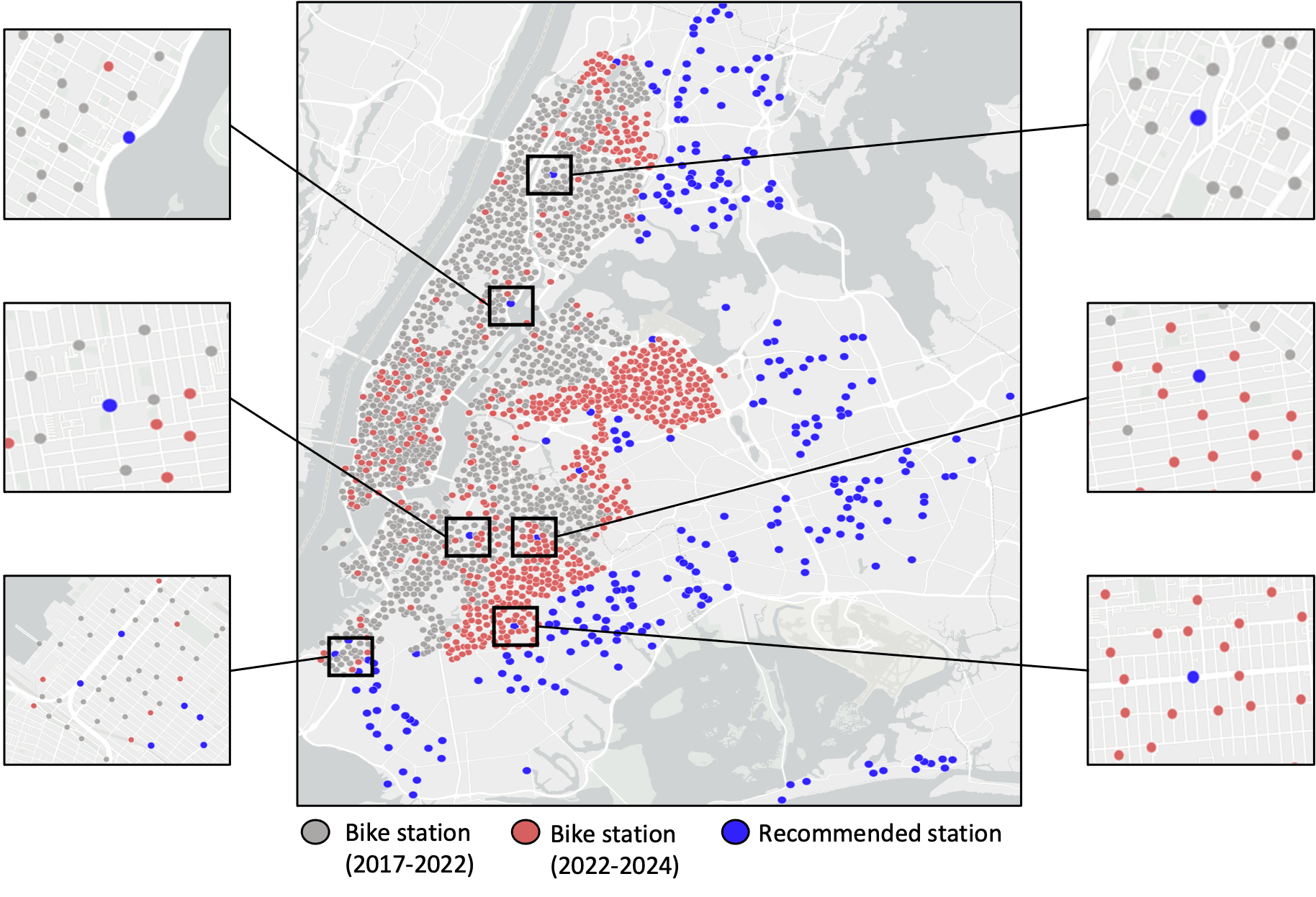}
    \caption{Spatial distribution of 280 recommended stations (blue), existing stations (gray), and previously planned stations (red). Station data is split into training (2017–2022) and testing (2022–2024) periods.}
    \label{analysis:rec_distribution}
\end{figure}

Furthermore, we visualize the spatial distribution of our recommended stations in Figure~\ref{analysis:rec_distribution}. The recommended stations (blue) effectively target underserved neighborhoods in the Bronx, eastern Queens, and southern Brooklyn, complementing existing (gray) and previously planned (red) Citi Bike stations. Our wAL-based site selection demonstrates the strategic future station recommendation to close gaps in the current BSS network, enhancing transit equity by extending accessibility to communities previously underserved by bike-sharing infrastructure. This highlights the effectiveness of our TFA framework in improving transport accessibility equity, aligned with SDG 11.2. 


%% file: section/conclusion.tex
\section{Conclusion}
In this study, we introduced \textbf{Transit for All (TFA)}, a spatial computing framework designed to support equitable expansion of BSS. TFA integrates multimodal geospatial data, including demographics, POI, and satellite imagery, to accurately predict bike demand at cold-start stations, significantly outperforming traditional approaches. 
Additionally, we developed a novel weighted Accessibility Level (wAL) metric, combining predicted bike demand with traditional accessibility measures. 

Our results showed that wAL effectively identifies previously overlooked disparities in transit accessibility among various income and racial groups, and provides strategic recommendations that enhance transit equity.
In the future, we plan to extend our framework to include diverse transportation modes and additional urban contexts, further supporting inclusive and adaptive mobility strategies aligned with the UN Sustainable Development Goal 11.2.